# Nirikshak: A Clustering Based Autonomous API Testing Framework

**Yash Mahalwal[1], Pawel Pratyush[1], Yogesh Poonia[1]**

[1]Department of Computer Science and Engineering, Maulana Azad National Institute of Technology, Bhopal 462003, India

## Abstract

Quality Assurance (QA) is a critical component in product development, particularly in software testing. Despite the evolution of automated methods, testing for REST APIs – a standard protocol for web services – often involves repetitive tasks. A significant portion of resources is dedicated more to scripting tests than to detecting and resolving actual software bugs. Additionally, conventional testing methods frequently struggle to adapt to software updates. However, with advancements in data science, a new paradigm is emerging: a self-reliant testing framework. This innovative approach minimizes the need for user intervention, achieving level 2 of autonomy in executing REST API testing procedures. It does so by employing a clustering method and analysis on logs categorizing test cases efficiently and thereby streamlining the testing process as well as ensuring more dynamic adaptability to software changes. Nirikshak is publicly available as an open-source software for the community at https://github.com/yashmahalwal/nirikshak.

**Keywords**- software testing, REST API, automation, log 0analysis, clustering

## 1. Introduction

Testing is crucial for production-grade software. It is the process of ensuring that the product is made correctly and meets all the expectations[1,2]. To achieve this, we need to understand all the specifications of the product and then form a set of validations. Subsequently, the product should be tested against the validations to discover any bugs. Understanding specifications and making validations needs human effort. After that, the product can be manually checked (manual testing) or we can write a script that does it for us (automated testing). With automated testing, QA is reduced to making appropriate validations[3]. These validations are written in the form of assertions. If the software fails an assertion, we discover a potential bug corresponding to it. It might also be possible that the product just differs from the specifications but the divergence is acceptable.

Now that we know about the process of software testing, it is clear that we need human intervention at two points:
1. Understand what is made and then decide what checks are needed to be performed.
2. Interpreting the results of validations.

An autonomous tool should take product specifications as its input. After that, it should be able to make proper validations. We already have automated ways to test software against a set of validations[1]. Our tool should then take the result of validations and provide us with appropriate insights. As with most artificial intelligence tasks, the mundane part of the process is the hardest to achieve. We need to be able to convey the specification to the tool. Moreover, the tool should be able to "see" the product thoroughly. In essence, we need a vision system. Once we have the results of validations, we have a plethora of techniques to extract insights from them. Most of the project is focused on making a vision system. After that, we have basic data analysis which proves the functionality of the vision system.

Autonomous testing seems like a fancy sci-fi concept. Needless to say, there is a lot of scope in the field[2,4,5]. Software testing is essentially understanding specifications and then validating the product against them. Most of the work can be automated with executable scripts. As discussed earlier in this section, human intervention is limited to a few fixed points.

### 1.1 Problems with automated testing

There are currently a few major challenges to most of the testing processes[4]. Major ones can be enumerated as follows:
1. Refactoring code requires refactoring all the test cases manually.

2. Code coverage is decided manually and is ultimately based on research done by a QA engineer.
3. As new features are added to the software, the older test cases have to be extended manually by the QA engineer. If that is not done, cases pertaining to functionality covered earlier will pass regardless of the impact of newer features - unless the new features interfere with existing functionality.
4. Eliminating bugs in the testing process itself is done manually by analyzing the relationship between the process entities. That can take up a considerable amount of time.
5. Since tests are written manually, there are always bugs in the testing process itself.

## *1.2 Advantages of artificial intelligence:*

Most of the drawbacks of traditional automated testing arise from the fact that there is significant involvement of the developer/tester in the process. AI effectively reduces these tasks[4,5,6]. As a consequence, the following things become possible:

1. *Reduced human intervention in tests*: A traditional testing tool will run the tests for us. But it would not know which tests to run to check a given part of the program. AI presents a scenario where the testing frameworks become 'intelligent' enough to analyze the current tests, changes in the code, code coverage, software usage, and traffic analytics and based on that, improve the test and run the tests as needed.
2. *Accelerating Manual Testing and Overall Process*: When software grows fast, the tests need to change equally fast in order to keep up with the pace. If not done, integrity, reliability, performance & security of software cannot be guaranteed. While changing the test cases manually is a tedious task, AI presents a scenario where all this can be done in a matter of seconds.
3. *Better test cases*: As changes are made to the software, the tester must analyze them to decide how current test cases should adapt. AI on the other hand, with solid data and facts, heuristically changes the cases as and when required. Since the tests are based on the tester's understanding of the business model, there are always scenarios where the tests may prove insufficient. AI reduces such cases by providing a more robust and powerful testing process.

## *1.3 Project scope*

There is quite a lot of work in autonomous user interface testing which as we already have complex vision systems for analyzing digital images[7]. Images are the perfect candidate as the data is properly structured and is easy to harness.

We move towards another field to prove the usability of our approach. One of the most popular programs in today's world is web APIs. Given their versatility and widespread usage, they form a good candidate for our project. The only problem here is that we need a structured model to be able to start with a vision system. Web APIs often follow well-defined architectures. Representational state transfer (REST) is one of the most popular web architectures[11]. Moreover, it provides enough constraints to give a well-structured model.

## *1.4 Measuring the abilities of the framework*

As Gil Tayar discusses in his famous medium article[8], autonomous testing is similar to autonomous driving. Autonomous testing tools can be classified based on levels of autonomy. Those levels are briefly explained below:

1. *Level 0 - No autonomy:* All the automation code is written by the QA engineer. This represents the traditional software testing approach.
2. *Level 1 - Drive assistance:* The better the vision system of the autonomous car is, the more autonomous it is. Similarly, the better AI can see the application, the more autonomous it can be. Once the testing framework can holistically see the application, it can help the QA tester write checks they usually write manually.
3. *Level 2 - Partial Automation:* Level 1 AI can detect changes in application with respect to a baseline i.e., how different it acts than expected. But it needs human intervention to review those changes. Figuring out whether the change is "good" or a bug is again a tedious task. Level 2 AI will be able to understand those changes and group changes throughout the application semantically. It will then be possible to accept or reject changes as a group.
4. *Level 3 - Conditional automation:* Level 2 AI can identify the changes by comparing the application with a baseline. It can analyze a given state of application but cannot understand if it is right or wrong. Level 3 AI goes a step beyond. It uses machine learning techniques to validate the application and figure out if the change is okay or if it is an anomaly. Human intervention is only needed when an anomaly is detected.
5. *Level 4 - High automation:* At Level 4, the AI will drive the testing process itself. It will be able to understand the application as a human does. It will do so by understanding user interactions and application structure. At that stage, it will be able to write and run tests and understand the results. It will use the level 2 and level 3 techniques to detect bugs.
6. *Level 5 - Full Automation:* This is complete automation. At this stage, AI will be able to converse with the project manager to understand the insight and drive the testing process. This clearly is fictional but gives a yardstick for an ideal tool.

## 2. Work Description

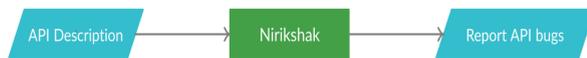

**Figure 3.1** Black box view

A higher-level view of the tool can be seen as above (Refer Fig 3.1). If we were to break it down, there are essentially the following components:
1. Vision system
2. Automated testing framework
3. Result analysis

Now we discuss the working of Nirikshak. We shall emphasize the functionality. Detailed information can be found on the official GitHub page[1].

### 2.1 Seeing the application

The REST semantics provide enough constraints over the architecture[9,10]. On top of that, we add two more simple constraints to provide a better structure. They are listed below:
1. We restrict ourselves to JSON as the engine of the application state.
2. Resources are independent.

These constraints allow us to take minimal information from the user and interpolate the rest of it. We have simple JSON schemas that help users describe the data models in the application. They map with real-world data types such as address, location, name, etc. Using the schemas, the user defines their resources and the endpoints.

Now that we have a way for users to describe the resources and endpoints with minimum effort, we have a way to understand the specifications. That is a core part of the vision. The next part is to be able to see the application. To achieve that, we have a very well-defined way of writing tests. We have established that the tool can understand API specifications. These tests help us validate against those tests thoroughly. The result of those tests helps us provide a view of the API.

To relate, we can think of each test case as a light ray directed towards the API (object). The ray hits the object and reflects back to us. If we capture the reflected ray, we can see the point where it hit the object. When the object is illuminated with sufficient light and viewed at the correct angles, it is possible to visualize it. In the same way:

---
1 https://github.com/yashmahalwal/nirikshak

1. Tests are a way to be able to see different parts of the application. A test can be thought of as a directed ray of light towards our object for the purpose of illuminating it.
2. Creating a test is not enough. After the ray of light reflects back, it is essential to capture and see the reflected ray. Similarly, the framework should see and understand the result of the test case.
3. If enough well-designed tests are used and the results correctly interpreted, we can get a view of the application.

### 2.2 Generating test cases

This section is concerned with the process of generating tests for a REST API such that they provide a complete view of the application. REST semantics are static and clear[9,10,11]. Therefore, they can be provided to the tool as procedural knowledge.

API endpoints are identified by the resource manipulation and the method used (an action that is taken). Each endpoint can be used in a variety of ways and every single way is a test case. To generate a test case, we need the following information:
1. **Endpoint description**: Resource, method, and user-defined input data templates.
2. **Expected behavior:** Decided by REST semantics and user-defined output templated.
3. **Resource instance for templates**: An appropriate resource instance to be used for populating templates. That is decided by the expected behavior.

#### 2.2.1 Case description

The user needs to provide the URL template of the endpoint and the method. Along with that, the user needs to provide templates for headers, query strings, and input body (if any). Based on the method, we know how the request tries to manipulate the API. Also, the expected behavior lets us know about the assertions we should make.

#### 2.2.2 Resource selection

Depending on the expected behavior, there are three categories of API requests[9]:
1. **Positive:** A valid request was made to the API that requested a valid operation. API was used as intended.
2. **Negative:** A valid request was made to the API which requested an illegal operation. The request was in violation of the API state.
3. **Destructive:** An invalid request was made to the API. This usually means that the request was made in the wrong way, such as sending an improper body.

A request is treated as an input to the API. The input along with the API state decides the outcome. The input format is fixed which is deduced using REST semantics and the description provided by the user. API state refers to the resources that are in the application. A collection of resources is maintained that is supposed to exist and use that to test our APIs. So it all boils down to selecting the correct resource to make the request.

1. **GET, DELETE, PATCH**: For positive requests, we generate request data using an existing resource. For negative requests, we use a non-existing resource.
2. **POST**: For positive requests, we generate request data using a non-existing resource. For negative requests, we use an existing resource.
3. **PUT**: We randomly select between an existing and a non-existing resource.
4. For destructive requests, we ask the user to provide us with an invalid format of request data. We then populate that format using an existing resource.

At this point, we know about the case we need to validate. We can also populate the data templated for the case using an appropriate resource instance. With these, we can write a test case.

## 2.3 Flow of testing

We can generate a test provided a scenario along with an appropriate resource instance. But it is essential to know what happens before all the tests and after them. For example, how do we know which resource instances exist in the API and which do not? Moreover, testing isolated endpoints does not map real word usage. We need to be able to mock complex flows with the API.

### 2.3.1 Setup operations

Before tests are run, we generate a collection of resources. The user hooks into this step and adds the resources to their application. At this step, we know about the resource instances that are supposed to exist in the application. The generated collection becomes our "local" collection of these resources. We define a *setupInstances* parameter that controls the number of generated resources. Since data is randomly generated, we provide an *iterations* parameter that repeats the testing process specified times to account for variation.

### 2.3.2 Request relationship graph

We have a hardcoded algorithm that reads the user's API description and then generates a graph of scenarios. We then traverse this graph, testing one scenario after the other. This mocks real-world usage of the API. We discuss this process in depth in section 2.4.

### 2.3.3 Generating and running tests

We have a generic test case template. At runtime, we link the template to user description files. The template uses these utilities to hook into the framework. These utilities allow the template to generate the request relationship graph and traverse it.

### 2.3.4 Result analysis

We use Jest[2], an automated testing tool. Our tests are written for Jest and are run using it. Jest generates well-defined structured logs. We encode case information into the logs. We then perform an analysis of the log data. We discuss this further in section 2.5.

### 2.3.5 Clean-up operations

After the tests are run, we allow the user to hook into the cleanup step. We provide the user with our local collection of resources. The user can remove them from the application at this point.

## 2.4 Components of the request relationship graph

As mentioned in section 2.3.2, the graph is generated using a hardcoded algorithm. The algorithm is trivial and can be found in the GitHub documentation[3]. The key part is traversing the graph generated.

### 2.4.1 Semantics of a request

A request is a combination of method, URL, body (if any), and headers. Currently, we support generating requests on the basis of these attributes. Future versions of the tool may have more advanced features such as supporting cookies. The key attributes of a request are its method and the outcome it is supposed to bring. Knowing these helps us know what the request is intended to do. We make assertions about the application and use requests to validate them. If the response from the API was as expected, our assertions are successful. Else they fail and so does the test.

### 2.4.2 Nodes

Based on user description, we collect each possible request scenario. This is simply a combination of request attributes discussed before. Each request is treated as a node of our graph. So a node is nothing but a combination of :

1. **Method**: Get/ Put/ Post/ Patch/ Delete
2. **Outcome case**: Positive/ Negative/ Destructive
3. **Method index**: The index of entry among similar methods. For example, if we have two ways to get a resource instance, we recognize them by giving them indices 0 and 1. This entry refers to that index. If there is only one entry for a method, the corresponding node has method index 0.
4. **Input data**: The data format to be used to hit the API. This includes headers, query string, and body (if any).
5. **Output data**: Data format of expected output from

---

2 https://jestjs.io/

3 https://github.com/yashmahalwal/nirikshak

the API. This includes headers, status, and body (if any).

Two nodes are considered equivalent if they make the same assertion. That means requests with the same method and outcome case are equivalent.

An example of a node is:
```
URL: /teacher/{resource:id}
METHOD: GET
TYPE: POSITIVE
INDEX: 0
```

Here, a `GET` request was made at the URL - `/teacher/<teacher-id>` and a Positive outcome is expected. This request is of index 0 meaning it is the first one of all equivalent `GET Positive` requests. If there is only one way to make a GET Positive request for the resource, only the $0^{th}$ index node exists. The data for this request is supposed to be populated using an existing student and the request expects the API to respond by indicating that the student exists.

### 2.4.3 Assertion result and API state

Assertions based on requests provide us with information about the application state. If we know that a given request succeeded (i.e., the corresponding assertion was validated), we can decide which requests should succeed from this point onwards. For every resource, we essentially maintain a graph of all the possible request sequences. Nodes of that graph are the requests that can be made (which are deduced from our API description). Edge from one node to another means that if the source node's assertion succeeds, we should expect the target node's assertion to succeed too.

### 2.4.4 Edges

Now we discuss the edges that can be in the graph. First of all, we generalize equivalent nodes. This means that all equivalent nodes have the same targets for outgoing edges and the same sources for the incoming edges. An edge between two nodes boils down to the method and the outcome case of the source and target nodes. We maintain a hardcoded algorithm that takes in two nodes and tells us if an edge is possible between them.

### 2.4.5 Traversal algorithm

This is the core of our test generation. We have a parameter called *steps*. We take the graph generated and traverse it in the following way:

1. Generate all walks of *steps* vertices from every node.
2. That is available as an array of strings. The inner array of strings is the nodes on a walk. The outer array is the collection of those walks.
3. We iterate over each walk. From every walk, we visit a node one by one. When we visit a node, we perform the appropriate testing actions.

The parameters affect the number of tests generated. Test cases grow exponentially with the number of steps and linearly with iterations. We recommend an upper bound of 3 steps and 5 iterations

### 2.5 Result analysis

We encode the graph node data into the test name. That data in turn is transmitted into test logs. Each test entry contains the following:

1. Test case outcome
2. Method - type and index
3. Resource name
4. URL
5. Iteration
6. Error message

We parse the log directly. Since the log was generated by us, we can provide certain guarantees regarding consistency and integrity. This trims a lot of preprocessing steps.

### 2.5.1 Test ratio

This is simply a ratio indicating the number of passed and failed test cases. It is represented by a donut chart.

### 2.5.2 Hierarchical grouping

We create a hierarchical distribution over the attributes represented by a hierarchical bar graph. We only perform grouping if there are any failed tests. This is simply grouping data into bins representing all permutations of test attributes.

### 2.5.3 Clustering

We perform clustering on five attributes (we do not consider iteration). These five attributes are test case outcome, method, resource name, URL, and error message. Since we do not have well-defined partitions, we perform a DBSCAN clustering [12,13]. The attributes are a mixture of nominal and ratio types. So we define custom distance functions for all the five attributes and a central distance function finds the weighted mean of these distances.

The implementation can be seen in the official documentation[4].

---
4 https://github.com/yashmahalwal/nirikshak

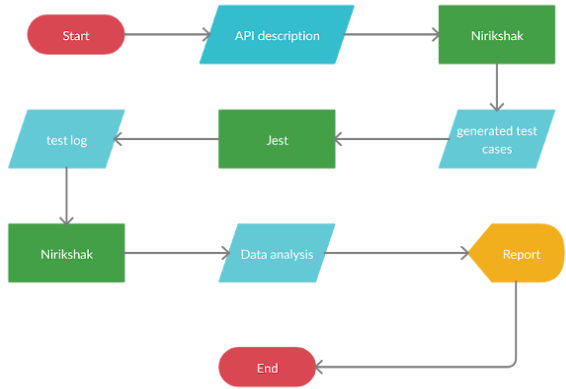

**Figure 3.2** A complete view of the tool

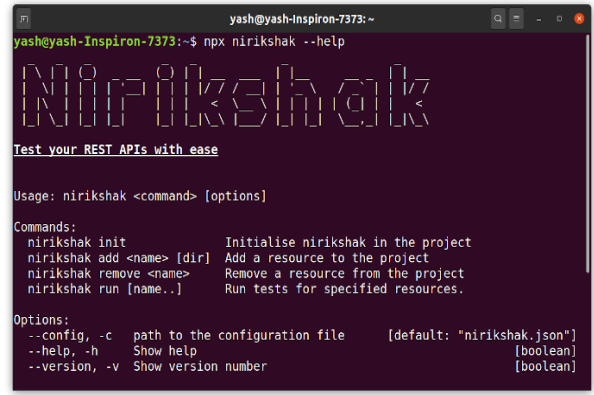

**Figure 5.1** A screenshot of the CLI

*2.5.4 Analysis steps*

The first step we take is to examine the number of total test cases and the number of failed tests. Based on these values, we proceed with our analysis. The following steps sum up the flow of analysis of our tool:

1. No test → No analysis
2. If any tests → Make a record of passed and failed tests (represented in the form of a donut chart)
3. If any failed tests → Perform hierarchical grouping of log data
4. If the number of failed tests exceeds a threshold of 100 → Perform clustering of the five attributes

# 3. Components

The tool is written in typescript and thoroughly tested. We expose three packages. Now we shall discuss each of them in the following subsections.

## 3.1 @nirikshak/core

This package exposes the core utilities for specification parsing and test case generation. Code for the package can be found at:

https://github.com/yashmahalwal/nirikshak/tree/dev/packages/core

## 3.2 @nirikshak/cli

This is a simple CLI for managing and running different test cases. This connects the @nirikshak/core package with Jest. Code for the package can be found at:

https://github.com/yashmahalwal/nirikshak/tree/dev/packages/cli

## 3.3 @nirikshak/reporter

This is a test reporter which hooks into Jest, collects the logs, and analyses them to form a report. Code for the package can be found at:

https://github.com/yashmahalwal/nirikshak/tree/dev/packages/reporter

## 3.4 Extending the test cases

If Nirikshak does not cover a certain test scenario, the user can manually enter the test case. We expose @nirikshak/core which provides utilities for writing tests manually.

# 3. Results

We ran the tool for a simple API that managed the data of students in a college. Below is the description of the API:

## 3.1 Resource description

The resource schema can be found at:

https://github.com/yashmahalwal/nirikshak/blob/dev/packages/simple-example/old/student/resource.json

## 3.2 Endpoints description

The Endpoint schema can be found at:

https://github.com/yashmahalwal/nirikshak/blob/dev/packages/simple-example/old/student/endpoints.json

## 3.3 Parameters

The values were set to the following:

1. *Steps* = 3
2. *Iterations* = 5
3. *Setup instances* = 10

To see the complete test results, view the animated results at:

https://github.com/yashmahalwal/nirikshak/blob/dev/docs/DataAnalysis.md

## 3.4 Test Ratio

With the bugs introduced manually, 318 tests failed and 488 passed.

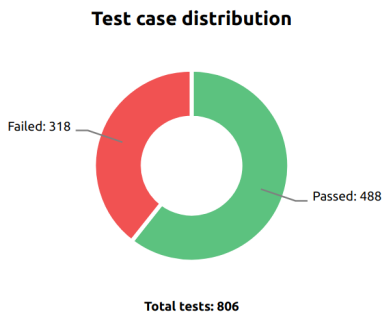

**Figure 6.1:** The Test Ratio

## 3.5 Hierarchical bar graph

We had granular information about the tests that passed and the tests that failed. We could find out about the categories and subcategories at all levels (Refer Fig 6.2).

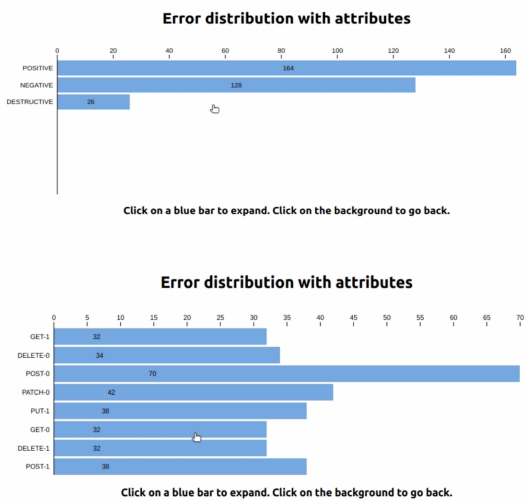

**Figure 6.2** A few hierarchies of the bar graph

## 3.6 Clustering

We performed a DBSCAN clustering with an epsilon value of 0.4 and minimum points of 7. That yields good results for repetitive points.

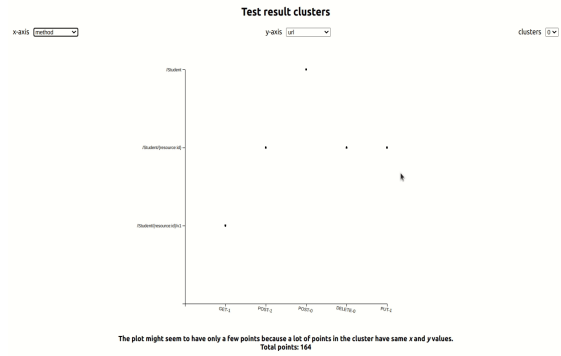

**Figure 6.3** Clusters formed

## 3.7 Capabilities of the tool

As per the model discussed in section 1.4, we have a **level 2** autonomous tool.

# 4. Conclusion and Future Scope

Given an API, a user needs to provide a very minimal description. With this, the tool has a powerful enough vision to visualize the API. The result of visualization is available as a fine-grained hierarchical bar graph. The tool generates a huge amount of test cases that query-string grows exponentially. The tool could successfully cluster the tool into groups with redundant points by balancing all the appropriate attributes. Now we discuss the future scope for the tool.

## 4.1 Schema processing

We can enhance user experience and tool capabilities by providing sophisticated parsers instead of simple parsing scripts. That can help us augment rich data models into the tool.

## 4.2 Testing provisions

We can provide more points for users to test the tool. For example, we can provide cookie and session support. We can also extend the framework to support non-functional tests like load testing, fuzz testing, etc.

## 4.3 Smarter data analysis

Currently, we provide basic data analysis as proof of the functionality of the vision system. Now that we have established that, we can move onto enhancing the data analysis process.

- **Level 3:** Classification of clusters as bugs or simple divergence.
- **Level 4:** Reinforcement learning to enhance the API description from application usage data.